\documentclass[aps,prstab,reprint,groupedaddress,floatfix]{revtex4-1}
\usepackage{amssymb, amsmath}
\usepackage{graphicx}
\usepackage{color}
\usepackage{xspace}
\usepackage{titlesec}

\newcommand{\degree}{\ensuremath{^\circ}\xspace}
\newcommand{\htp}{\ensuremath{\mathrm{H}_2^+}\xspace}
\newcommand{\BE}[0]{\begin{equation}}
\newcommand{\EE}[0]{\end{equation}}
\newcommand{\BEA}[0]{\begin{eqnarray}}
\newcommand{\EEA}[0]{\end{eqnarray}}

\newcommand{\nuebar}{\ensuremath{\bar{\nu}_e}\xspace}

\mathchardef\mhyphen="2D

\newcommand{\figref}[1]{Figure~\ref{#1}}
\newcommand{\tabref}[1]{Table~\ref{#1}}

\newcommand{\DD}{DAE$\delta$ALUS\xspace}

\titleformat{\section}
{\center\normalsize\bfseries\uppercase}{\thesection}{1em}{}

\titlespacing\section{0pt}{12pt plus 4pt minus 2pt}{6pt plus 4pt minus 2pt}
\titlespacing\subsection{0pt}{12pt plus 4pt minus 2pt}{6pt plus 4pt minus 2pt}
\titlespacing\subsubsection{0pt}{12pt plus 4pt minus 2pt}{6pt plus 4pt minus 2pt}

\begin{document}
\title{Update on the IsoDAR 60~MeV/amu Cyclotron Acceleration Simulations}
\author{Daniel Winklehner}
\email{winklehn@mit.edu}
\affiliation{for the \DD Collaboration}
\date{\today}

\begin{abstract}
In this technical report, the work done by members of the \DD collaboration
on simulating the acceleration of 5 mA of \htp 
in the IsoDAR compact isochronous cyclotron is summarized. The findings are that sufficient turn separation between the $(\mathrm{N}-1)^{th}$ and $\mathrm{N}^{th}$ turn can be achieved with
either initially matched but also with mismatched beams by careful placement of collimators 
to scrape away halo particles before the beam energy has reached 2 MeV/amu. Here we define
"sufficient" as $<200$ W beam loss on the septum.
In the acceleration simulations, beam losses on the collimators vary from 10\% to 25\%
depending on the initial beam conditions. 
The collimator placement in the central region, halo formation, and turn separation at the 
final turn are discussed. Finally, a scheme to further mitigate the risk of beam loss on the
extraction septum by placing a narrow stripping foil in front of it is investigated. We also 
present the findings of a separate study subcontracted to the company AIMA, 
who investigated injection of \htp ions through a spiral inflector and accelerating them 
for four turns. Here the reported beam loss before 2 MeV/amu was 58\%, but bunches were 
not accelerated further, so we are not reporting this as a final result.
\end{abstract}

\pacs{}

\maketitle

\section{Introduction}
A 10~mA 60~MeV/amu cyclotron would have enormous impact in neutrino physics through 
the IsoDAR project \cite{bungau:isodar,adelmann:isodar, abs:isodar}, for isotope 
production \cite{schmor:isotopes, alonso:isotopes, waites:isotopes}, and as a 
pre-accelerator for a 10~mA, 800~MeV - 1~GeV cyclotron that can 
be used for ADS(R) \cite{ishi:adsr,rubbia:adsr, biarrotte:ads,lisowski:ads} 
and particle physics, including the \DD experiment 
\cite{alonso:daedalus, aberle:daedalus, abs:daedalus, calabretta:daedalus, conrad:daedalus}.
Potential uses are summarized in \tabref{tab:uses}.
Because this concept originated within the development of \DD, 
this multi-use 60~MeV/amu cyclotron is 
historically called the \DD Injector Cyclotron, or DIC.
Here we use the terms \emph{IsoDAR cyclotron} and \emph{DIC} interchangeably.

Among the challenges for the DIC are the strong space charge effects of 
such a high intensity beam and the small phase acceptance window of the isochronous 
cyclotron.\ Space charge matters most in the Low 
Energy Beam Transport Line (LEBT) and during injection into the cyclotron.
Our concept to mitigate these risks is based on three novelties:
1. Accelerating 5 mA of \htp instead of 10 mA of protons, leading to the same number of 
nucleons on target at half the electrical current as the remaining electron bound in 
the \htp molecular ion reduces the electrical current in the beam.
2. Injecting into the compact cyclotron via an RFQ partially embedded in the yoke to
aggressively pre-bunch the beam. 3. Designing the cyclotron main acceleration to optimally
utilize vortex motion. 
In the following subsections, the injection and central region design will be briefly summarized.
In Section \ref{sec:main}, the previously published and new unpublished work by collaboration members 
will be presented and reviewed. Finally, the implications for IsoDAR and further mitigation
methods for beam loss on the extraction septum, using a stripping foil placed right before
the septum will be discussed in Section \ref{sec:discussion}.
\begin{table*}[t]
    \setlength{\tabcolsep}{4pt}
	\vspace{-10pt}
	\footnotesize
	\caption{A few potential uses for high current proton beams and how cyclotrons can be 
	         leveraged to reach the goals. 
             ADSR: Accelerator Driven Sub-critical Reactors, 
             ADS: Accelerator Driven Systems for nuclear waste transmutation.
             Cyclotrons can be a cost-effective alternative for tests and demonstrations at the
             low-power end of the spectrum (tens of mA).
             Adapted from~\cite{winklehner:nima}.}
	\label{tab:uses}
	\centering
    \vspace{5pt}
    \renewcommand{\arraystretch}{1.25}
		\begin{tabular}{lllll}
            \hline
            \textbf{Application} & \textbf{Field} & \textbf{Current}  
                                 & \textbf{Energy} & \textbf{Comment}\\
            \hline \hline
            IsoDAR \cite{bungau:isodar, adelmann:isodar, abs:isodar} & 
            neutrinos & 
            10~mA & 
            60~MeV & 
            Use \nuebar from decay-at-rest to search for sterile neutrinos.\\
            
            \DD \cite{alonso:daedalus, aberle:daedalus, abs:daedalus,
                      calabretta:daedalus, conrad:daedalus} & 
            neutrinos & 
            10~mA & 
            800~MeV & 
            A proposed search for CP violation in the neutrino sector.\\
            
            ADSR \cite{ishi:adsr,rubbia:adsr} & 
            energy & 
            10-40~mA & 
            $\sim 1$~GeV & 
            Cyclotrons are a cost-effective alternative for demonstrator experiments.\\
            
            ADS \cite{biarrotte:ads,lisowski:ads} & 
            energy & 
            4-120~mA & 
            $\sim 1$~GeV & 
            Cyclotrons are a cost-effective alternative for demonstrator experiments.\\
            
            Isotopes \cite{schmor:isotopes, alonso:isotopes, waites:isotopes} & 
            medicine & 
            $1-10$~mA & 
            3-70~MeV & 
            E.g.: 10 mA/60 MeV can increase worldwide $^{225}$Ac production by 6000 \cite{alonso:isotopes}.\\
            
            Material testing & 
            fusion & 
            10-100~mA & 
            5-40 MeV & 
            Testing of fusion materials similar to IFMIF \cite{moeslang:ifmif}, at lower power.\\
            \hline
		\end{tabular}
\end{table*}

\subsection{Injection}
\begin{figure}[t]
    \centering
    \includegraphics[width=0.5\columnwidth]{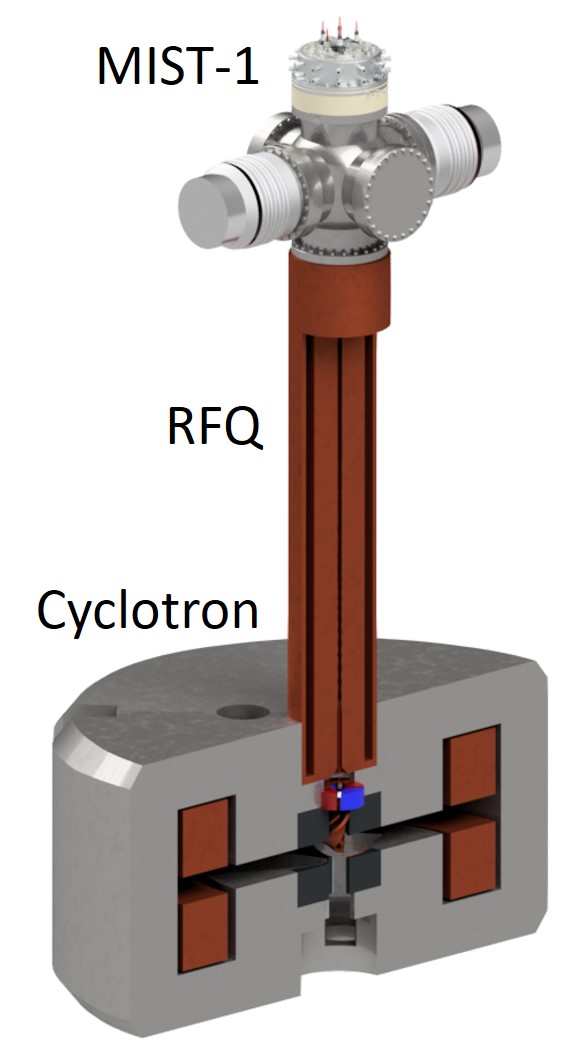}
    \caption{Cartoon rendering of the RFQ Direct Injection Prototype. Ions are produced in
             the ion source (top), are accelerated and bunched in the RFQ (middle) and 
             injected into the cyclotron central region to be accelerated to 1 MeV/amu
             (bottom).}
    \label{fig:rfq-dip}
\end{figure} 
In order to accelerate \htp, it has to be produced in an ion source in sufficient 
quantities. At MIT, we have built a prototype ion source optimized for \htp production: MIST-1.
For a description, see our recent publication on high intensity cyclotrons 
\cite{winklehner:nima} and references therein.
Systematic studies are ongoing and a publication
about the performance of the source is forthcoming.
The ion source design is based on a previous multicusp ion source developed at LBNL, 
which reported a ratio of 20\% protons and 80\% \htp, as well as a maximum beam 
density of 50 mA/cm$^2$ \cite{ehlers:multicusp1}.
The maximum beam density from MIST-1 so far was $40$ mA/cm$^2$, and the best \htp fraction
was seen at $25$ mA/cm$^2$, with 20\% p$^+$, 20\% \htp, 50\% $\mathrm{H}_3^+$ and 10\% 
$\mathrm{H}_2\mathrm{O}^+$ (unpublished). As commissioning of MIST-1 is ongoing, we are using 
the 80\% \htp reported in \cite{ehlers:multicusp1} for all simulations.
In Ref. \cite{winklehner:nima}, we also describe the physics design of an RFQ linear accelerator-buncher 
which is going to be embedded in the cyclotron yoke and will deliver a highly bunched beam
to the spiral inflector -- an electrostatic device that bends the beam from the axial direction
into the acceleration plane of the cyclotron (median plane, or mid-plane), where the beam
is accelerated and matched to the cyclotron main acceleration (described in Section \ref{sec:main}). 
A prototype of this \emph{RFQ Direct Injection Project (RFQ-DIP)} is currently under construction
(see \figref{fig:rfq-dip} for a cartoon rendering). By aggressively pre-bunching the beam, 
we fit more particles into the RF phase acceptance window of $\approx20\degree$.
Due to the high bunching factor and strong space charge, the beam starts diverging in transverse
direction and de-bunching in longitudinal direction soon after the exit of the RFQ.
To mitigate this, a re-bunching cell has been included in the RFQ design
and an electrostatic quadrupole focusing element has been placed before the 
spiral inflector. 
\begin{figure}[b]
    \centering
    \includegraphics[height=0.6\columnwidth]{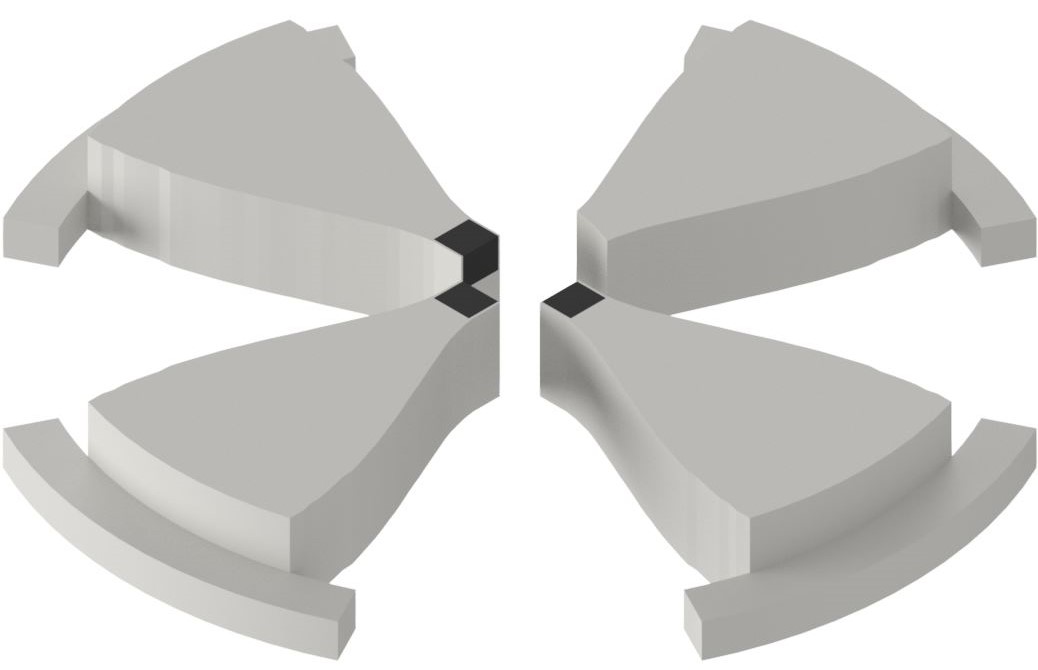}
    \caption{CAD rendering of the iron poles for magnetic field calculations (lower half only). 
             The vanadium-permendur inserts can be seen in black at the pole tips. 
             One of the pole tips is truncated, yielding space for the spiral inflector.}
    \label{fig:aima_poles}
\end{figure} 
In addition, the spiral inflector electrodes can be carefully shaped to add vertical 
focusing as well. Transmission from the ion source to the exit of the spiral inflector
was $\approx78\%$ for both test cases (10 mA and 20 mA of total beam current, 
20\% protons, 80\% \htp) with transverse emittances of $0.3-0.4$ mm-mrad (1-RMS, normalized) 
and $7-8$ keV/amu-ns longitudinal emittance (1-RMS)~\cite{winklehner:nima}.

\subsection{Central region}
\begin{figure}[b]
    \centering
    \includegraphics[height=0.6\columnwidth]{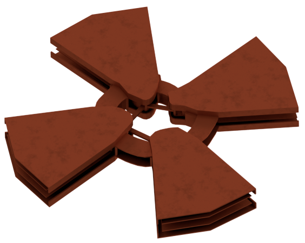}
    \caption{CAD rendering of the dee-dummydee electrodes, posts and additional grounded
             elements, used to calculate the electric fields in the cyclotron 
             central region.}
    \label{fig:aima_electrodes}
\end{figure}

Particle distributions obtained in the injection simulations were used in a detailed 
central region study subcontracted to the company AIMA Developpement in France and
comulated in a technical report \cite{aima:central}. In this study, a 3D magnetic field
was generated that includes the effects of vanadium-permendur (VP) inserts in the 
pole tips and one poletip cut short to allow placement of the spiral inflector
(see \figref{fig:aima_poles}). 
The VP has a sharper turn in the B-H curve, slightly improving the flutter in the
central region. An optimized dee electrode system was generated during the study,
which can be seen in \figref{fig:aima_electrodes}. This system exhibits good vertical focusing
and small orbit center precession. The dee peak voltage was increased to 80 kV from the 
nominal 70 kV in the IsoDAR baseline, which is still tolerable. 
\begin{figure}[t]
    \centering
    \includegraphics[width=\columnwidth]{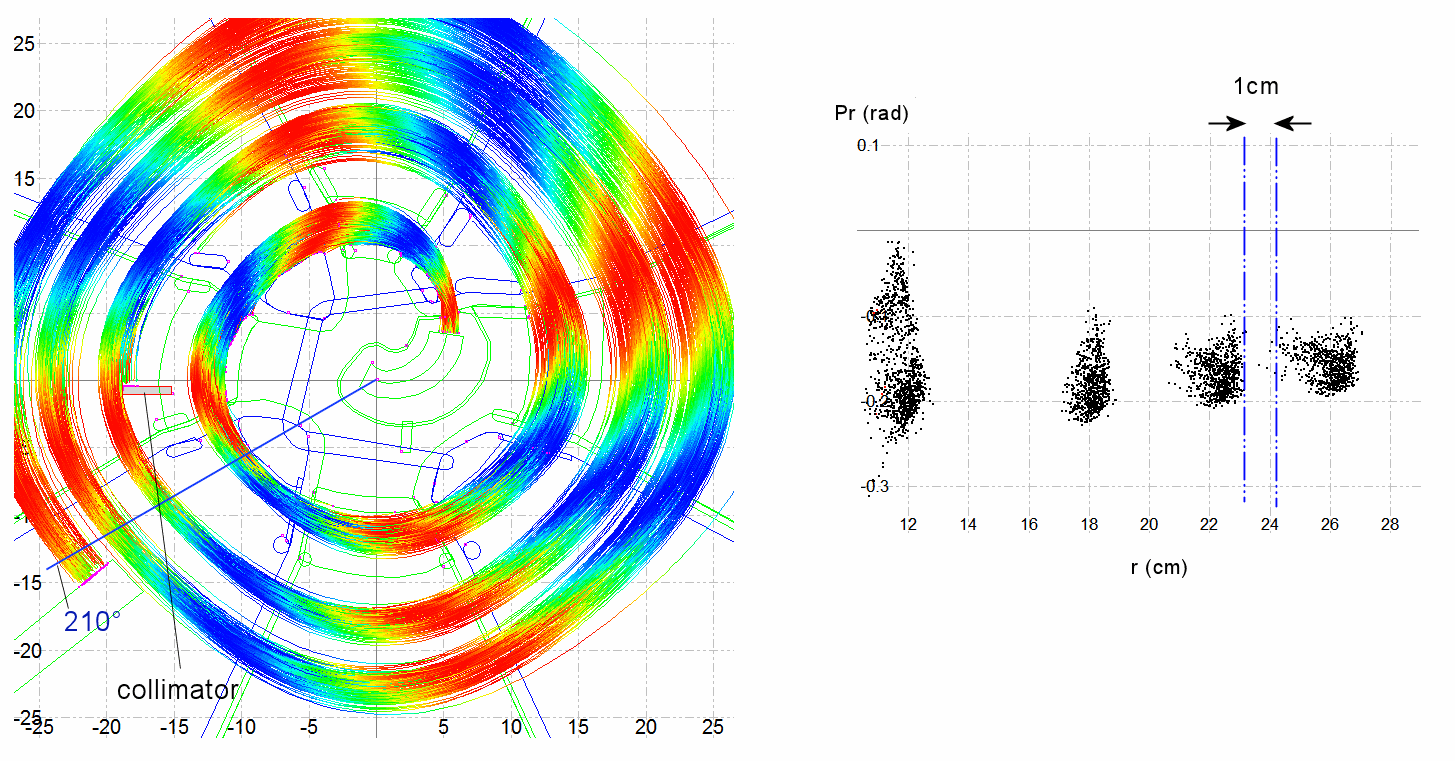}
    \caption{Left: Trajectories of the first 3.5 turns (2 MeV) in the simulated central region.
             Right: Demonstrated turn separation of 1 cm (edge-to-edge) after placing a collimator
             in the first turn. Beam transmission from the entrance of the spiral inflector
             to the probe was 42\%. From \cite{aima:central}.}
    \label{fig:aima_result}
\end{figure} 
By placing a single collimator in the first turn,
the desired edge-to-edge turn separation of 1 cm was achieved during the fourth turn 
(at 1 MeV/amu beam energy). This led to a beam loss of 58\% in spiral inflector and 
central region and a cumulative transmission efficiency of 42\%. 
Note that this replaces the 78\% presented in the previous subsection as AIMA used 
their own spiral inflector model.
As no additional significant losses are anticipated in acceleration and extraction,
in order to obtain 5 mA of beam, about 12 mA \htp need to be injected into the RFQ.
Assuming 80\% \htp, we are looking at a total extracted current of 15 mA. RFQ simulations
have shown that we can get good results up to 20 mA with the current design \cite{winklehner:nima}. 
However, this central region study, as of yet, does not include space 
charge effects (space charge was included up to the entrance of the spiral inflector).
Inclusion of space charge will change the beam dynamics (vortex-effect).
Furthermore, collimator placement is restricted to one location. 
Other studies (see Section \ref{sec:main}) suggest that better results can be obtained 
with multiple collimators in the first 3-4 turns.
Both, space charge and collimator placement, will be investigated further.

\section{Simulations of the IsoDAR main acceleration}
\label{sec:main}
\begin{figure}[t]
    \centering
    \includegraphics[width=0.8\columnwidth]{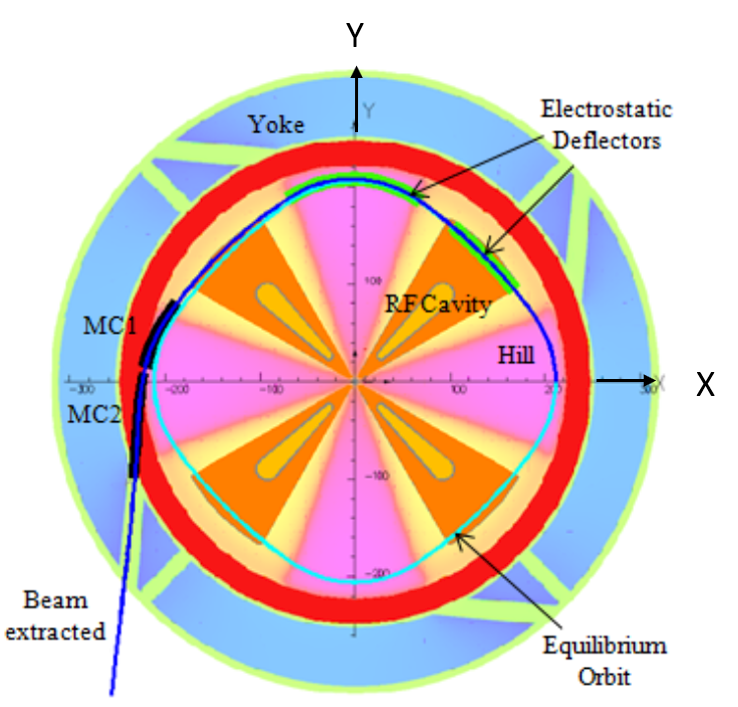}
    \caption{Cartoon of the IsoDAR cyclotron. Indicated are the hills (magenta) and valleys (yellow) 
             of the isochronous field, the four double-gap RF cavities
             (centered around 45\degree, 135\degree, 225\degree, and 315\degree), 
             the 60 MeV/amu static equilibrium orbit,
             and examples of deflector and magnetic channel (MC1 and MC2) placements.
             The outer radius of the yoke is 3.2 m.
             From \protect\cite{calanna:isodar}.}
    \label{fig:calanna_cycl}
\end{figure}

\begin{figure}[b]
    \centering
    \includegraphics[width=1.0\columnwidth]{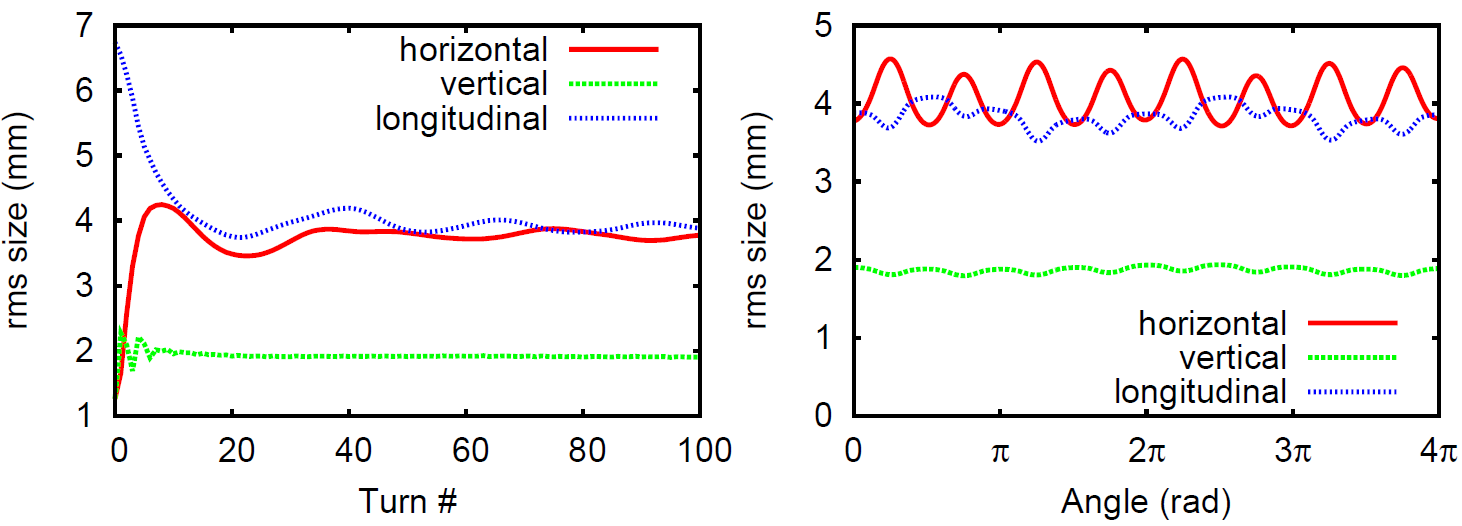}
    \caption{The rms size snapshot at 0\degree azimuth in 100 turns (left), 
             and the rms envelope in the last two turns (right), for
             a coasting (unaccelerated) beam with 5 mA and 1.5 MeV/amu. This demonstrates the
             formation of a stationary (matched) distribution from an initially
             unmatched beam.
             From \protect\cite{yang:daedalus}.}
    \label{fig:yang_coasting}
\end{figure}

The simulations of the IsoDAR main acceleration were done in three steps: In 2013, 
Jianjun Yang performed a set of studies for the 60 MeV/amu \DD Injector Cyclotron (DIC), 
which later became the IsoDAR cyclotron, and the 800 MeV/amu \DD Superconducting
Ring Cyclotron (DSRC), demonstrating that with careful collimator placement and by 
including neighboring bunch effects across several turns, beam loss on a 0.5 mm thick 
extraction septum, placed at R = 200 cm, could be kept below the required 
200 W \cite{yang:daedalus}. 
The starting condition was a 1.5 MeV/amu beam on the fourth turn. 
In 2016, Jakob Jonnerby extended these studies down to starting energies 
of 193 keV/amu (first turn) for his master's thesis \cite{jonnerby:thesis}. 
In 2017, Maria Yampolskaya further refined the collimator placement during a summer 
internship at MIT. 
An important factor in the clean extraction is the vortex effect, a combination of 
external focusing forces in the isochronous cyclotron and space-charge forces from the
Coulomb interaction of the particles in the high intensity beam \cite{baumgarten:vortex1}. 
Vortex motion has first been seen in PSI Injector II and subsequently reproduced with 
simulations in a previous publication \cite{yang:vortex}.
All simulations in this section were performed using the well-established parallel particle-in-cell
code OPAL that takes into account space charge \cite{adelmann:opal}. A top-down view
of the IsoDAR cyclotron is shown in \figref{fig:calanna_cycl}. In harmonic mode 6, the opening
angle of the cavities was chosen to be 28\degree. Example placements of electrostatic deflectors
(with septum) can be seen as well. All azimuthal angles in this section are measured from -180\degree to
180\degree, with 0\degree the positive x-axis.

\subsection{Early simulations starting at 1.5 MeV/amu}
\begin{figure}[t]
    \centering
    \includegraphics[width=1.0\columnwidth]{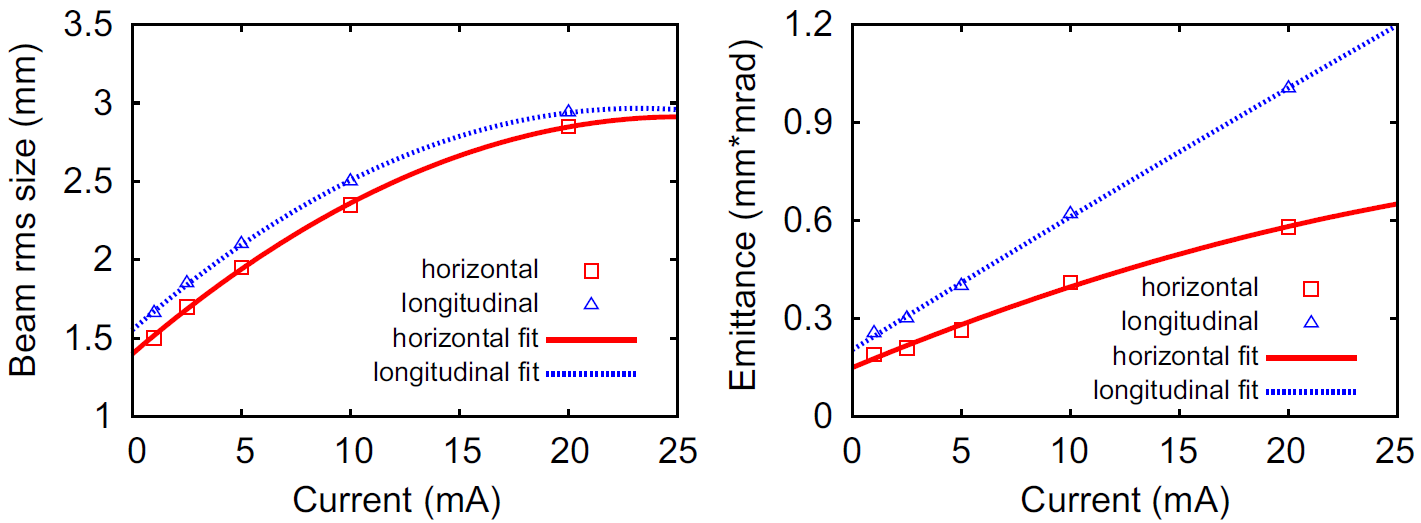}
    \caption{Stationary rms beam size (left) and normalized rms emittance (right) 
             vs. beam current.
             From \protect\cite{yang:daedalus}.}
    \label{fig:yang_matched_size}
\end{figure}
In this subsection, Jianjun's work from 2013 \cite{yang:daedalus} is summarized.

\subsubsection{Matched coasting beam}
To demonstrate the formation of a stable matched distribution, 
a coasting beam was first tracked for 100 turns without acceleration in
the IsoDAR cyclotron nominal magnetic field.
The initial beam was Gaussian and had rms beam sizes of 1.26 mm in transverse
and 6.7 mm in longitudinal direction (corresponding to 40\degree full phase width).
In \figref{fig:yang_coasting}, the results are shown, demonstrating the formation 
of a matched beam of 4 mm rms size within $\approx 20$ turns. The beam size then oscillates
with the number of sectors (\figref{fig:yang_coasting}, right). Stationary rms beam size 
and normalized rms emittance are plotted vs different beam currents in \figref{fig:yang_matched_size}.
It can be concluded that no flattop cavity is necessary and all four 
valleys can be used for accelerating dee-dummydee structures, yielding high energy
gain per turn and hence higher turn separation at extraction.

\subsubsection{Accelerated beam}
\begin{figure}[b]
    \centering
    \includegraphics[width=0.8\columnwidth]{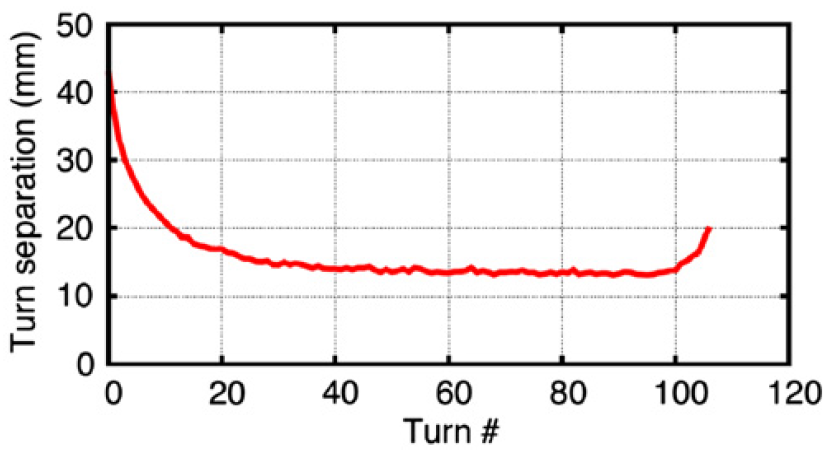}
    \caption{Center-to-center turn separation as a function of turn number. 
             From \protect\cite{yang:daedalus}.}
    \label{fig:yang_turnsep}
\end{figure}
\begin{figure}[t]
    \centering
    \includegraphics[width=1.0\columnwidth]{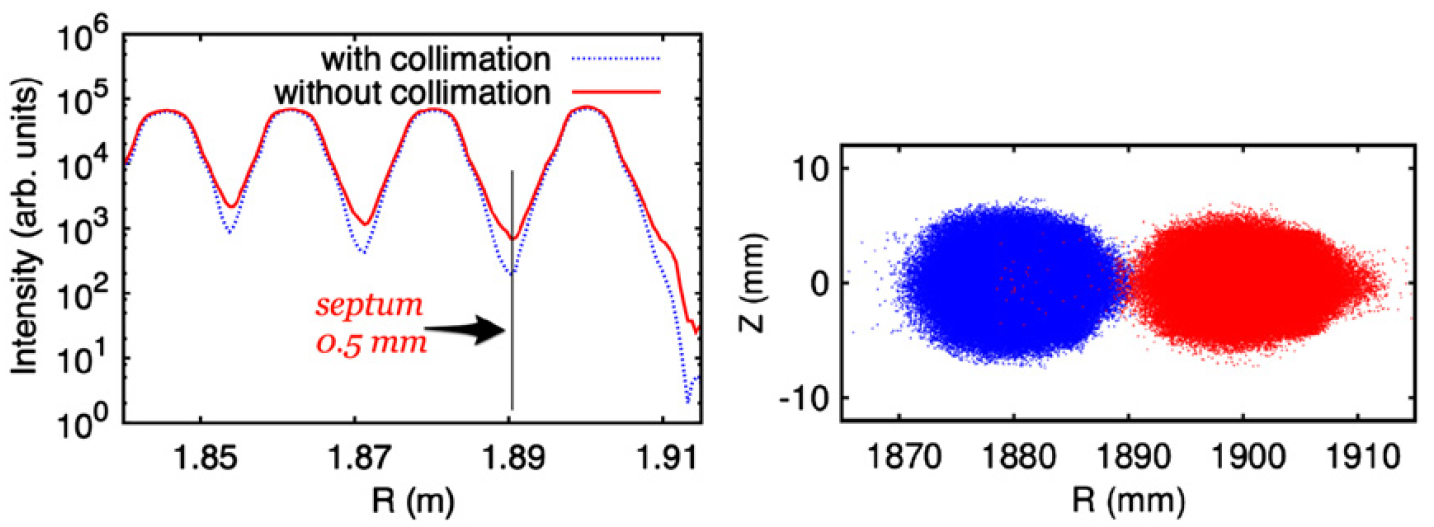}
    \caption{The radial profile of the last four turns at the center of a valley 
             (left) and the r–z projection of the collimated particles during the last 
             2 turns (right) for 5mA of beam current. 
             The total macroparticle number in a bunch is $10^6$. 
             From \protect\cite{yang:daedalus}.}
    \label{fig:yang_final_turns}
\end{figure}
The acceleration simulations in \cite{yang:daedalus}
start at the exit of the central region. Considering that both the
space charge effects in the injection line and the transverse–
longitudinal coupling motion in the spiral inflector inevitably
increase the emittance, the initial normalized emittance at the
exit of central region is set to a large value compared to typical
ion sources, i.e. 0.6 $\pi$-mm-mrad. The phase acceptance and initial
energy spread are assumed to be 10\degree – 20\degree and 0.4\% respectively. In
order to reduce the tail particles of the extracted beam, four
collimators are placed at around 1.9 MeV/amu to cut off about
10\% of the halo particles. High turn separation during the final turns is 
achieved by a combination of the vortex effect, using the $\nu_r = 1$ resonance (sharp
radial field drop), 
and a high energy gain per turn. In~\cite{yang:daedalus}, the separation
of the final turn is $\approx 20$ mm for a 5 mA beam (see \figref{fig:yang_turnsep}).
The turn separation is shown in \figref{fig:yang_final_turns} and the loss on the septum is
listed in \tabref{tab:yang_losses} for 1, 5, and 10 mA using 10\degree and 20\degree initial
phase. 
\begin{table}[b]
	\vspace{-10pt}
	\footnotesize
	\caption{Beam loss power on the 0.5 mm wide septum for 90\% duty cycle. 
	         From \protect\cite{yang:daedalus}.}
	\label{tab:yang_losses}
	\centering
    \vspace{5pt}
    \renewcommand{\arraystretch}{1.25}
		\begin{tabular}{llll}
            \hline
            \textbf{Injection phase width} & \textbf{1 mA} & \textbf{5 mA}  & \textbf{10 mA}\\
            \hline \hline
            10\degree & 27 W & 108 W & 1080 W \\
            20\degree & 63 W & 99 W & 1422 W \\
            \hline
		\end{tabular}
\end{table}
These simulations, while using an initial beam with higher than usual emittance and phase 
width to provide a safety margin, neglect the energy spread introduced by the spiral inflector, 
and start with a matched distribution at 0.54 MeV/amu. Ideally, the results of the central region
study would be used for the main acceleration simulations. This is a work in progress,
currently ongoing at MIT. However, one can also close in on the matching point of central
region and main acceleration by extending the latter to lower starting energies and refining 
the collimator placement. This is reported in the following subsection.

\subsection{Extension of OPAL simulations to lower starting energies}
\begin{figure}[t]
    \centering
    \includegraphics[width=0.9\columnwidth]{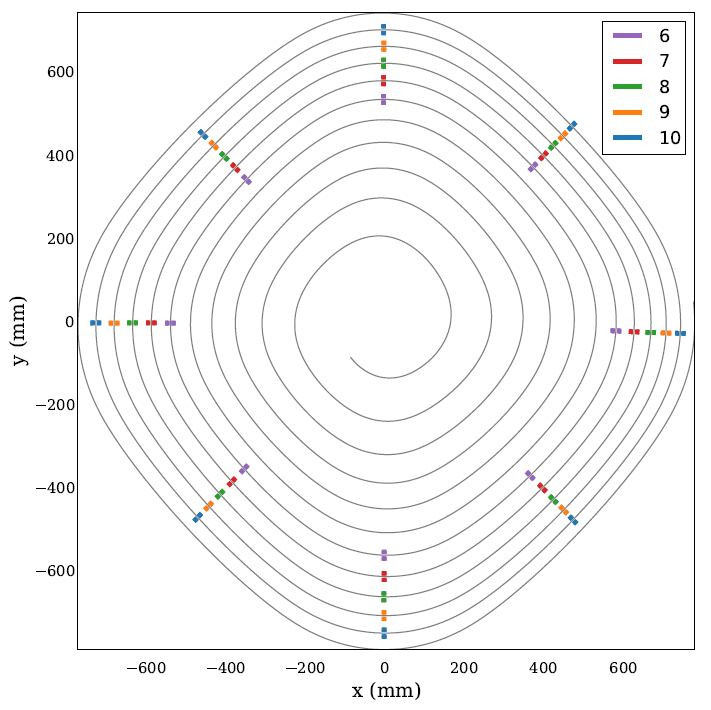}
    \caption{Possible placements of collimators in turns 6 - 10 overlaid with the 
             design trajectory from single particle tracking. Collimators were only placed
             in a single turn for each simulation.
             From \protect\cite{jonnerby:thesis}.}
    \label{fig:jonnerby_collimators}
\end{figure}
\begin{figure}[b]
    \centering
    \includegraphics[width=0.9\columnwidth]{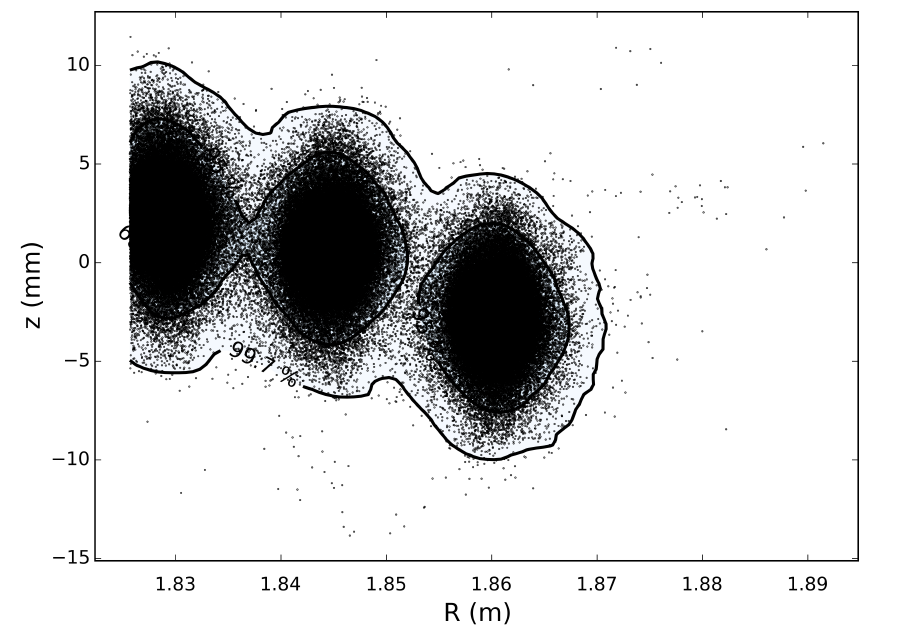}
    \caption{The two last turns projected on the r -- z plane, at an azimuthal angle of -45\degree
             in the cyclotron.
             From \protect\cite{jonnerby:thesis}.}
    \label{fig:jonnerby_final_turns}
\end{figure}

In his ETHZ master's thesis \cite{jonnerby:thesis}, Jakob Jonnerby extended the
IsoDAR cyclotron simulations to a lower starting energy of 193 keV/amu and refined
the collimator placement. He went through a similar process as described in the 
previous subsection in order to find a matched distribution. The studied collimator 
placements are shown in \figref{fig:jonnerby_collimators}. The best results were 
obtained with collimators placed either in turn 8, 9 or 10. With the lower 
starting energy and different collimator placement, Jonnerby reports a turn separation
at 60 MeV/amu of 15 mm center-to-center (which is lower than the 20 mm of Yang) and
states that the beam power loss on the septum is larger than the limit of 200 W. 
The r–z projection of the last two turns at an azimuthal angle of -45\degree is 
shown in \figref{fig:jonnerby_final_turns}. A major difference between Jonnerby's and 
Yang's work is that Jonnerby ends the simulations at r = 1.86 m and Yang at 
$\mathrm{r}  = 1.9 \,\mathrm{m}$. 
\begin{figure}[t]
    \centering
    \includegraphics[width=1.0\columnwidth]{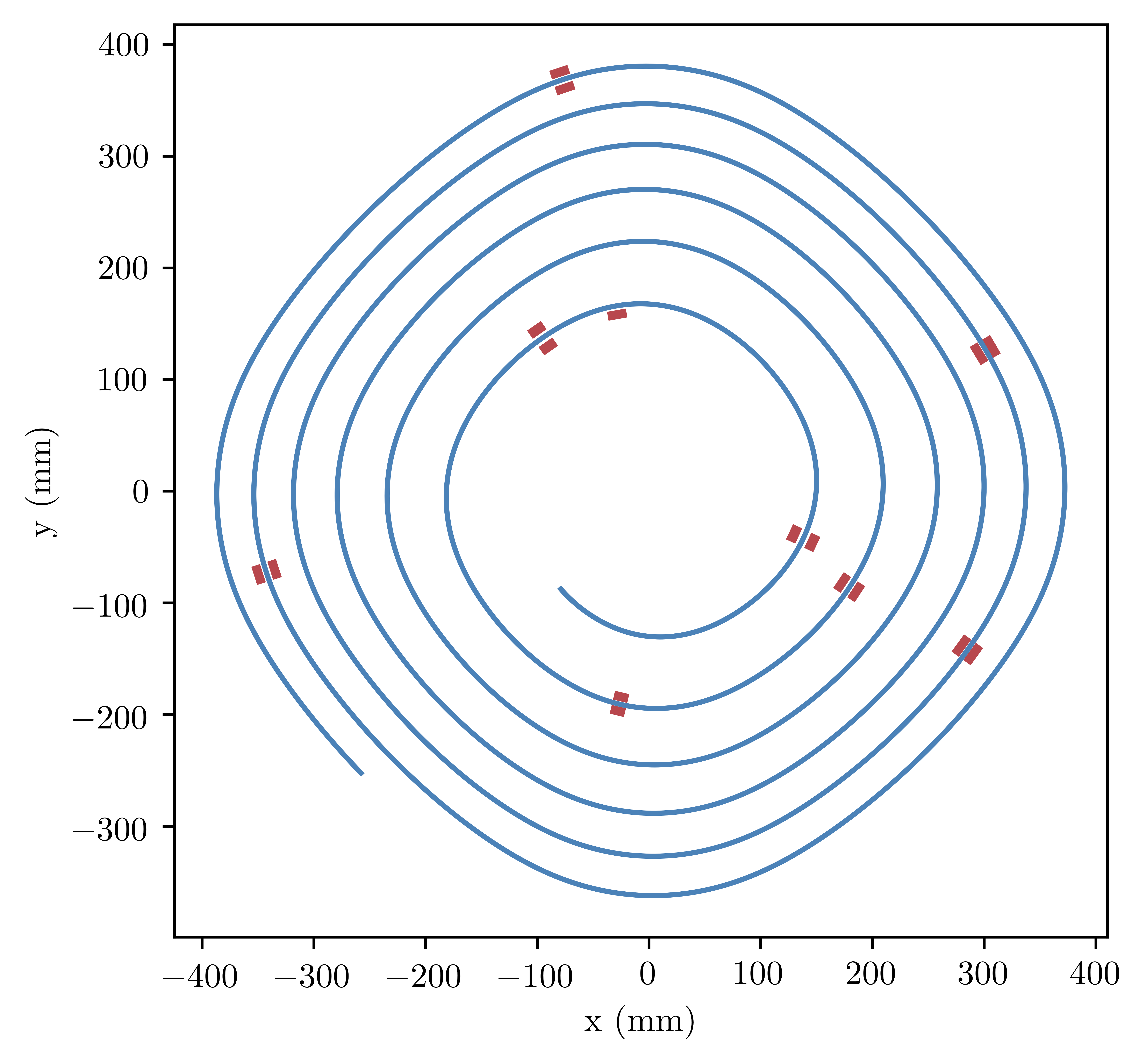}
    \caption{The bunch centroid trajectory of the first 6 turns (blue) with 9 collimators (red).}
    \label{fig:yampolskaya:coll1}
\end{figure}
\begin{figure}[b]
    \centering
    \includegraphics[width=1.0\columnwidth]{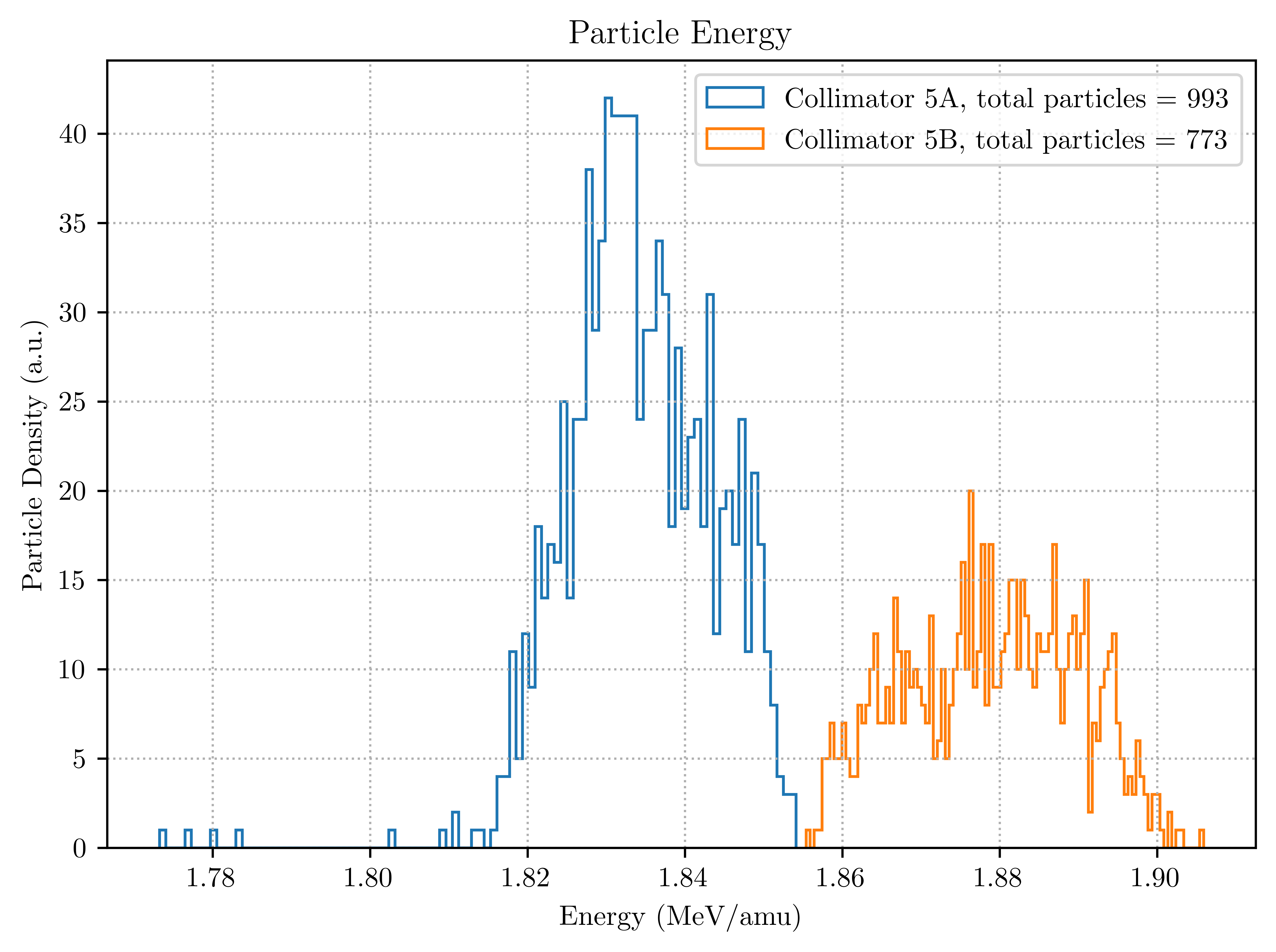}
    \caption{Energy histogram of particles lost on collimator 9. Total number of accelerated particles = 100,000.
             Collimator 5A is on the inside of the orbit, Collimator 5B on the outside.}
    \label{fig:yampolskaya:coll2}
\end{figure}
Tracking for more turns, while increasing the beam energy beyond 60 MeV/amu, would
bring the beam closer to the $\nu_r=1$ resonance region of the magnetic field, further
increasing the separation at the expense of beam quality.
It was shown in his work, that a matched distribution can be found at lower energy and
that the beam exhibits vortex motion, stabilizing the beam growth, leading to 
similar beam sizes as reported by Yang. The beam loss on the collimators is
$\approx 23\%$. Maria Yampolskaya, during a summer internship at MIT in 2017, 
refined Jonnerby's collimator placement and was able to improve the beam 
quality and turn separation close to extraction. DW, for this report, re-ran 
several of Yampolskaya's simulations, tuned the RF phases slightly, and 
further optimized the collimator placement. This work is, as of yet, unpublished.

\begin{figure}[t]
    \centering
    \includegraphics[width=1.0\columnwidth]{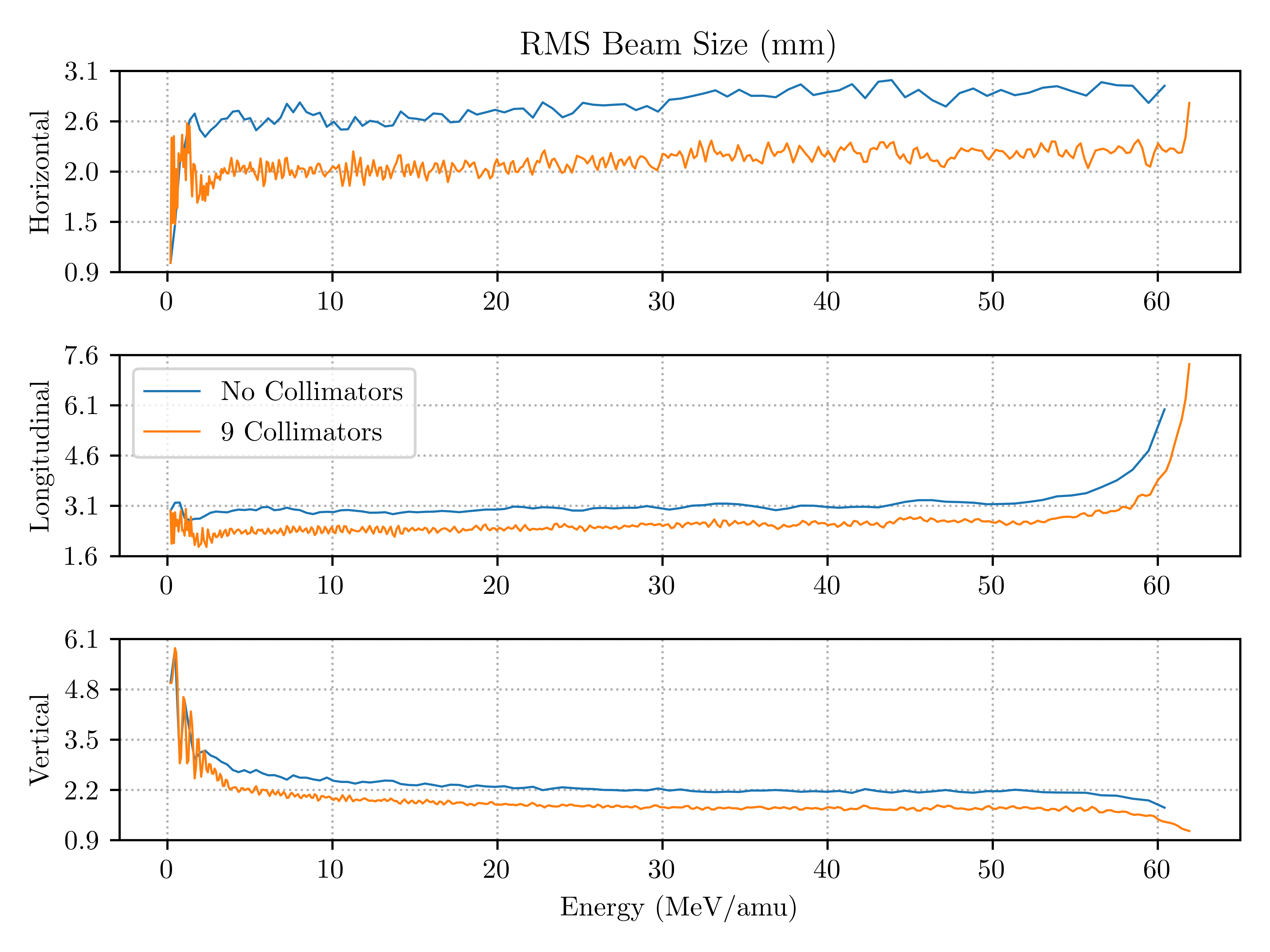}
    \caption{RMS beam size for two cases: No collimators and 9 collimators. A clear reduction can be seen 
             with collimators. Also visible is the effect of the $\nu_r=1$ resonance above 60~MeV/amu.
             The longitudinal and radial beam size are approximately the same above 5~MeV/amu, due
             to the vortex effect.}
    \label{fig:yampolskaya:rms}
\end{figure}
\begin{figure}[t]
    \centering
    \includegraphics[width=1.0\columnwidth]{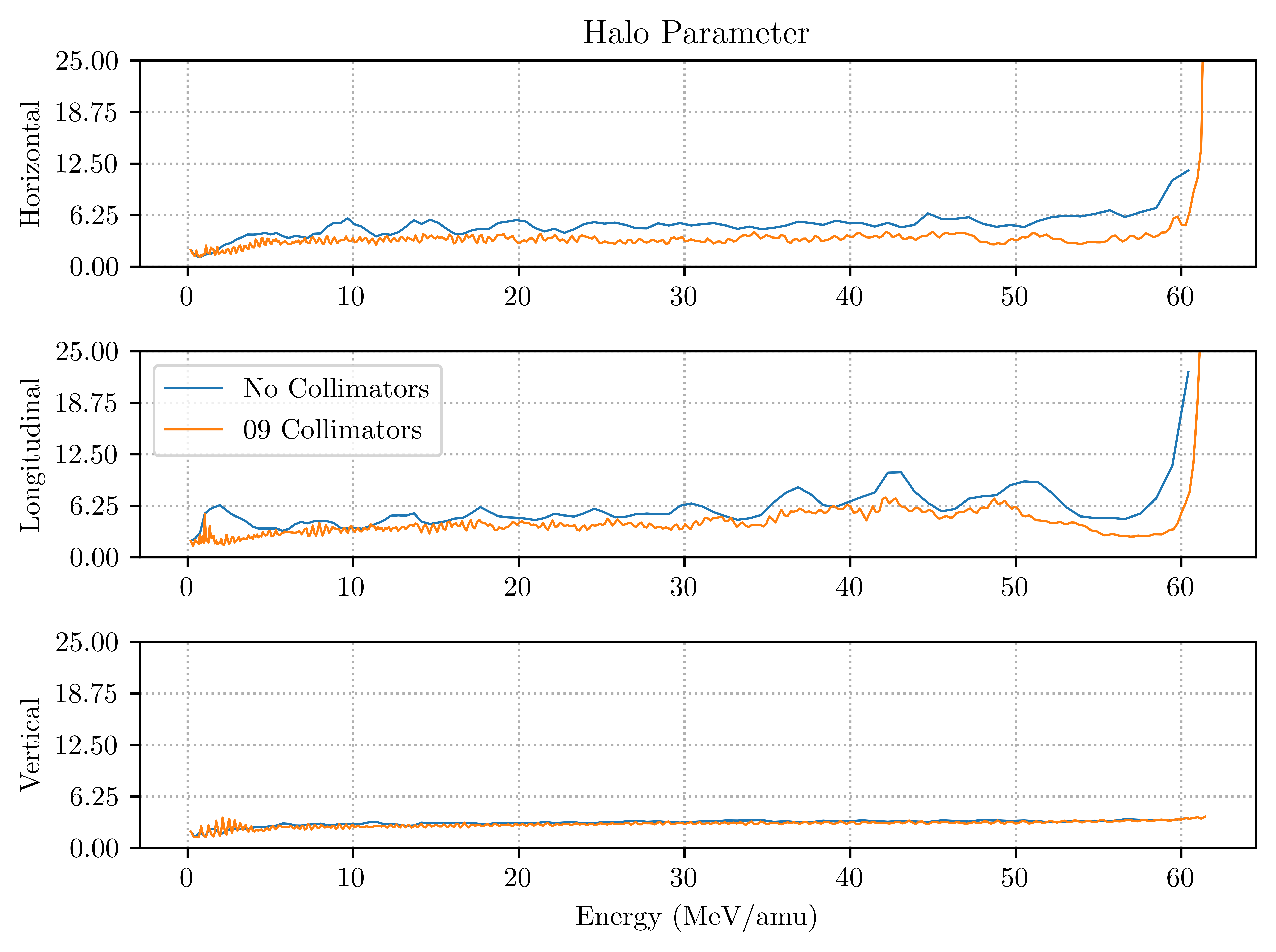}
    \caption{Halo parameter (cf. \eqref{eq:halo}) for two cases: No collimators and 9 collimators. 
             A clear reduction can be seen with collimators. Also visible is the effect
             of the $\nu_r=1$ resonance above 60~MeV/amu.}
    \label{fig:yampolskaya:halo}
\end{figure}

The final result for simulations in harmonic mode 6, starting with a mean beam energy of 193 keV/amu
and placing the bunch on a matched orbit at -135\degree yielded deposited beam power on
the septum of 98~W. For this, 9 collimators were placed in the first 6 turns (see \figref{fig:yampolskaya:coll1}). 
The highest energy particles lost on the collimators were $\approx1.9$~MeV/amu (cf. \figref{fig:yampolskaya:coll2}), 
similar to Yang's choice. 
The RMS beam size for a case without collimators and with the 9 collimators as well as the halo parameter,
defined as:
\BE
\label{eq:halo}
H = \frac{\langle x^4 \rangle}{\langle x^2 \rangle^2}-1
\EE
are shown in \figref{fig:yampolskaya:rms} and \figref{fig:yampolskaya:halo}, respectively.
A clear reduction of both can be seen with collimators. Also visible is the effect
of the $\nu_r=1$ resonance above 60~MeV/amu. The overall beam loss in the central region
in this simulation was 27\%.
\begin{figure}[b]
    \centering
    \includegraphics[width=1.0\columnwidth]{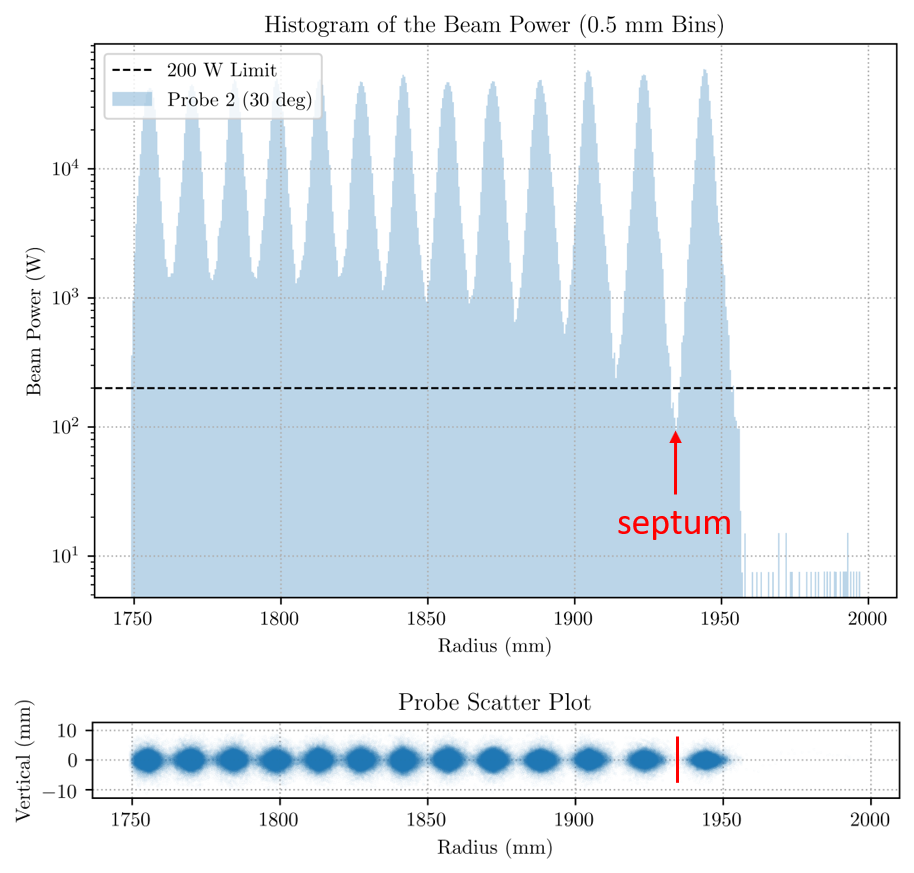}
    \caption{Probe 2 (30\degree). The beam power binned in 0.5 mm bins (this is a conservative choice
             for septum width) versus radius (top). R-Z scatter plot of particles passing through
             the probe (bottom). It can be seen that a septum inserted at the right radial position
             wuld only take about 98 W of power.}
    \label{fig:yampolskaya:probe}
\end{figure}
The deposited power on a probe, located at 30\degree azimuth, 
sorted in 0.5 mm bins (a conservative choice for septum thickness), is shown
in \figref{fig:yampolskaya:probe} (top) with an R-Z scatter plot of the particles passing through the probe
(bottom). Septum placement is indicated in red. A few stray particles with higher energies are still visible, 
which will deposit their energy in other parts of the cyclotron, adding to activation. These are few, 
and can also be suppressed with further collimator optimization.
These results show clearly that, even with a beam starting at 193 keV/amu (essentially during the innermost turn),
collimators can be placed such that the turn separation at extraction is sufficient, albeit with beam losses around 25\% in the central region. 

\section{Discussion}
\label{sec:discussion}
\begin{figure}[t]
    \centering
    \includegraphics[width=1.0\columnwidth]{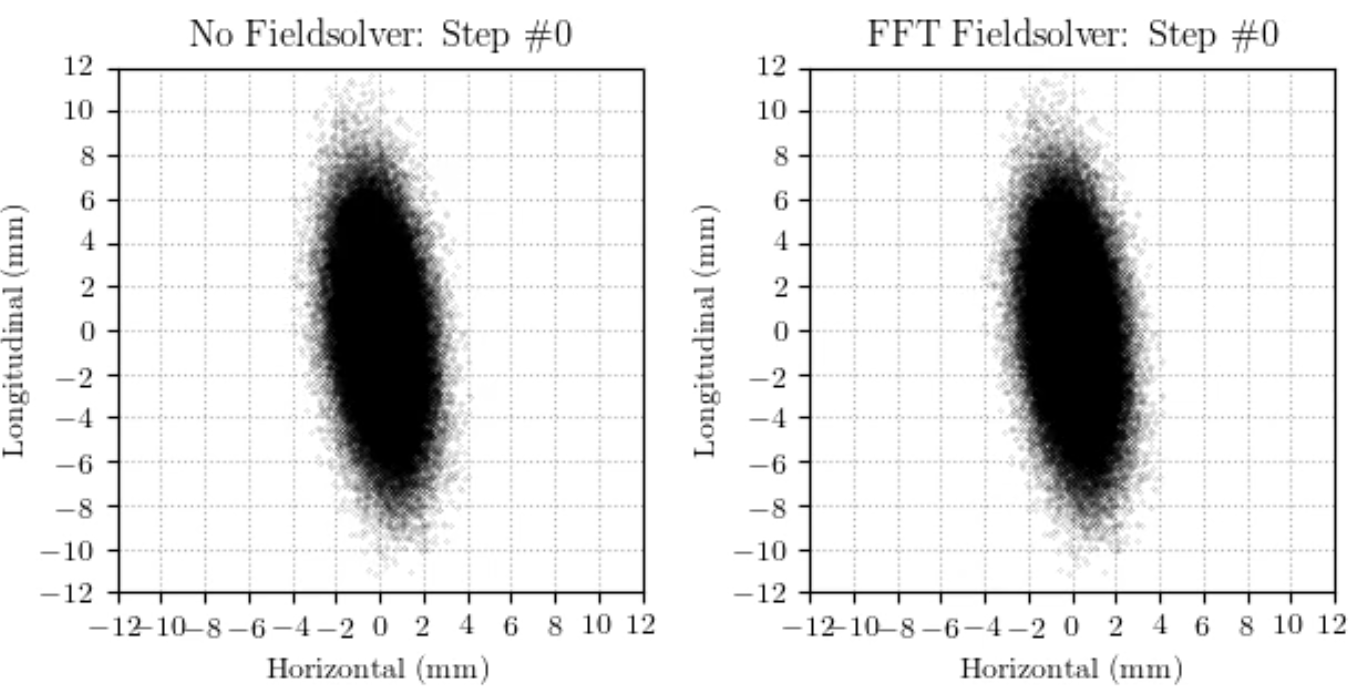}
    \includegraphics[width=1.0\columnwidth]{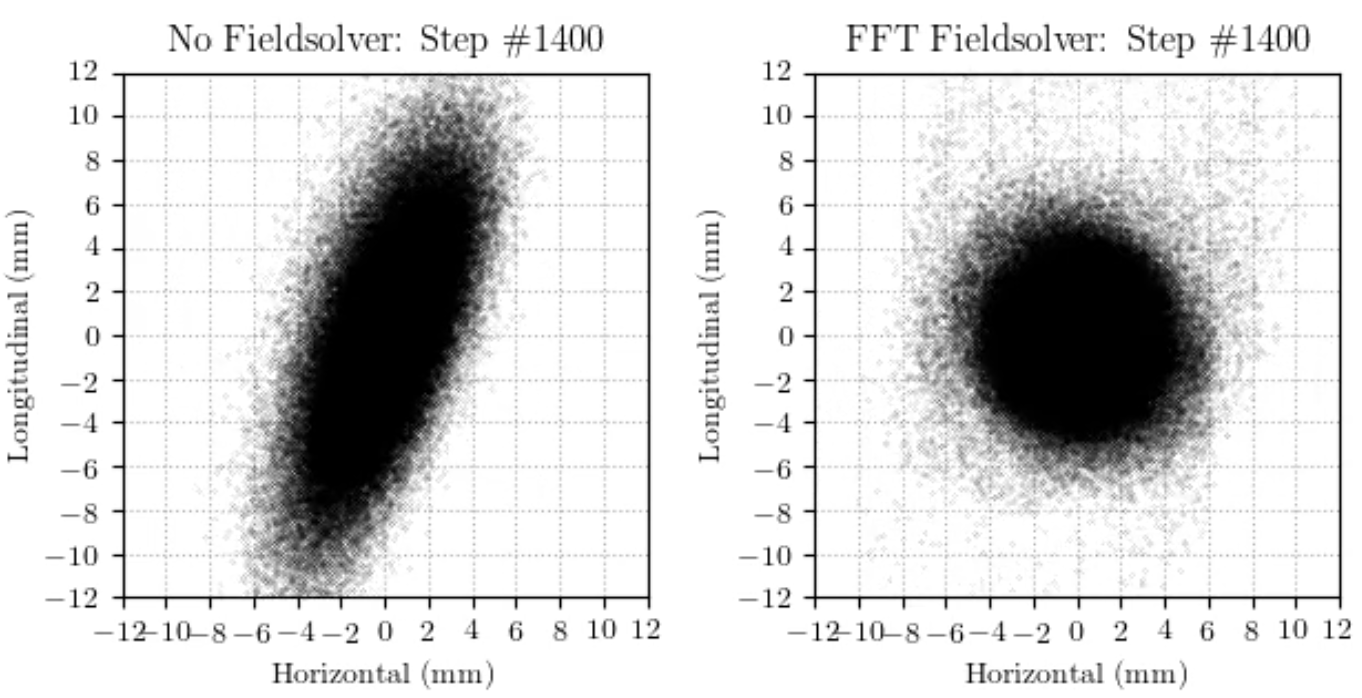}
    \caption{Top: 
             Starting bunch in local frame ('longitudinal' is the direction 
             of the bunch mean momentum).
             Bottom: 
             Bunches in local frame after 7 turns. 
             No space-charge (left), space-charge for 6.65 mA (right).
             By coupling space-charge and external focusing, a round distribution with
             halo develops. Identical initial bunches were used.}
    \label{fig:yampolskaya:bunches}
\end{figure}

All 60 MeV/amu simulations point to a similar conclusion: The beam exhibits vortex motion, which
generates a stable round distribution. At the same time, a halo is formed that needs to be
removed by placing collimators early-on. One interesting observation is that, starting at 193 keV/amu,
it seems to take $\sim$ 4-6 turns for this process to reach a stage where the beam is matched and 
halo removal becomes efficient (i.e. no new halo is generated). One can see an example of the bunch
in local frame (longitudinal = direction of mean momentum, transverse $\sim$ radial direction, 
both centered at 0) in \figref{fig:yampolskaya:bunches}.
There seems thus to be a balance between 
removing particles before their energy becomes high enough to activate parts of the cyclotron and
late enough for vortex motion to have produced a matched distribution.

Another important point is the effect of the $\nu_r = 1$ resonance on the beam size. It can be 
seen in \figref{fig:yampolskaya:rms} and \figref{fig:yampolskaya:halo} that after 60 MeV/amu, the 
longitudinal beam size and halo parameter both increase exponentially. At the same time,
the beam precession induced by the resonance is what leads to the good turn separation in the last turn.
In the next step, a septum and appropriate electric field will be modelled  and beam will be guided to,
or even through a simple extraction channel to see how the beam quality (size, emittance, energy spread) 
changes during the extraction process. 

\subsection{Protecting the septum with a stripping foil}
\begin{figure}[b]
    \centering
    \includegraphics[width=1.0\columnwidth]{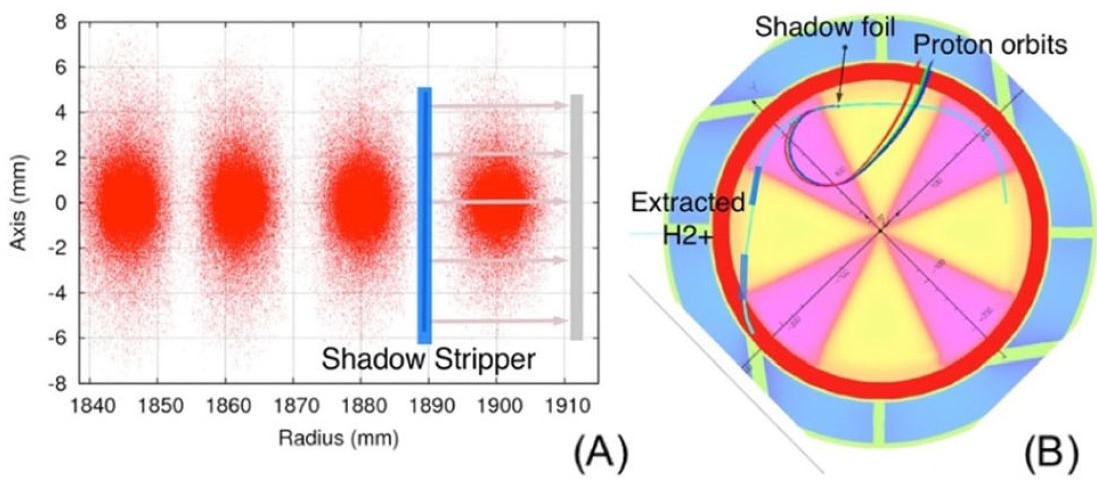}
    \caption{(A) Placing a thin carbon foil intercepts particles that would otherwise hit the 
             septum. These \htp ions are stripped of their electron and the
             resulting protons (B) are extracted from the cyclotron and safely terminate
             in a beam dump. From \cite{waites:isotopes}.}
    \label{fig:waites_stripping}
\end{figure}

To mitigate beam loss on the septum, a new concept was introduced by collaborator
Luciano Calabretta and has been discussed in the context of medical isotope 
production in \cite{waites:isotopes}. 
The septum can be protected by placing a shadow foil in front of it. This carbon foil intercepts
particles that would otherwise hit the 
septum. \htp ions are stripped of their electron and the
resulting protons are extracted from the cyclotron and safely terminate
in a beam dump. We estimate that about $\sim 0.02\%$ of the beam will be lost in this way,
a number that is insignificant compared to the beam loss in the central region.
The scheme is illustrated in \figref{fig:waites_stripping}.

\section{Conclusion}
In this report, we summarized the IsoDAR 60~MeV/amu simulations using the
particle-in-cell code OPAL. Early on, it was shown by Jianjun Yang that a beam, matched or mismatched,
will eventually, through vortex motion, become approximately round ($\mathrm{x}_{\mathrm{rms}} \approx \mathrm{y}_{\mathrm{rms}}$). By placing collimators during the first several turns of acceleration,
halo developing in the matching process can be removed in a controlled manner without 
activation of the machine. These simulations were then repeated and extended 
from a starting energy of 1.5~MeV/amu down to to 0.193~MeV/amu (essentially the first turn) 
with similar results. Beam loss on these collimators varies between 10\% and 25\%,
with power deposition on the septum around 100~W (half of the safety limit of 200~W
determined at PSI).
These are high fidelity simulations, using 1e5 to 1e6 particles per bunch. 
However, further sensitivity studies are still necessary to determine how the 
system will react to variations in input beam current and placement, as well as 
additional energy spread from the spiral inflector. In the meantime, we propose
a mitigation method in which \htp particles which would hit the septum are stripped of their
remaining electron by means of a narrow foil. The resulting protons have reduced
magnetic rigidity and will go on a different path, missing the septum. They have to
be extracted from the cyclotron and safely dumped (or used for a different purpose).

\subsection{Outlook}
All simulations of the IsoDAR cyclotron in this 
technical report were performed with harmonic 6. In the recent past, however,
the collaboration has made the decision to switch to harmonic 4 and the 
simulations are currently repeated with the lower frequency. Preliminary results
with a wider cavity opening angle of 42\degree show very similar results to the ones
presented here for harmonic 6. This report will be updated in the near future to
include the new results. The same holds for multi-bunch simulations, wherein OPAL
injects five bunches in sequence (one per full turn for five turns) to account for the
space charge effect of neighboring bunches. The results are not dramatically changed.
The final step will be to repeat the AIMA central region simulations with space-charge
and use the output particle distribution as input in the OPAL IsoDAR 60~MeV/amu
simulations.

\section{Acknowledgments}
This work has been supported by the Heising-Simons Foundation, 
and the US National Science Foundation under grants NSF-PA-1912764 and NSF-MRI-1626069.

\bibliography{bibliography}

\end{document}